\documentclass[onecolumn,showpacs]{revtex4}
\usepackage{amssymb}
\usepackage{makeidx}
\usepackage{amsmath}
\usepackage{graphicx}
\usepackage{epstopdf}
\usepackage{dcolumn}
\usepackage{bm}

\input{tcilatex}
\begin{document}

\title{Induced by coherent THz-radiation high harmonics generation in
bilayer graphene at high Fermi energies}
\author{A.G. Ghazaryan $^{a} $}
\email{amarkos@ysu.am}
\author{H.H. Matevosyan $^{b} $}
\author{Kh.V. Sedrakian$^{a} $}
\affiliation{$^{a} $ Centre of Strong Fields Physics, Yerevan State University, 1 A. Manukian,
Yerevan 0025, Armenia\\
$^{b} $ Institute Radiophysics and Electronics NAS RA, 1 Alikhanian brs., Ashtarak 0203, Armenia\\}
\date{\today }

\begin{abstract}
The higher order harmonic generation process in the nonperturbative regime
at the interaction of coherent electromagnetic radiation with the AB-stacked
bilayer graphene at high Fermi energies is considered. The applied coherent
low-frequency radiation field in the high Fermi energy zone of electrons
excludes the interband transitions enhancing high harmonic rates. The
developed microscopic nonlinear quantum theory for charged carriers
interaction with a strong pump wave is valid near the Dirac points of the
Brillouin zone. The Liouville-von Neumann equation for the density matrix in
the multiphoton excitation regime is solved both analytically and
numerically. Based on the numerical solutions, we examine the rates of
higher-order harmonics of the pump wave of arbitrary polarization. Obtained
results show that bilayer graphene can serve as an effective material for
the generation of higher order harmonics from THz to the mid-IR domain of
frequencies at the pump wave moderate intensities.
\end{abstract}

\pacs{78.67.-n, 72.20.Ht, 42.65.Ky, 42.50.Hz, 32.80.Wr, 31.15.-p}
\maketitle



\section{Introduction}

The nonperturbative regime of emission of harmonics - high harmonic
generation (HHG) \cite{1Corkum}, \cite{hhg1} is one of the main nonlinear
phenomena at the interaction of strong coherent electromagnetic (EM)
radiation with the matter \cite{Abook}, which has been widely studied both
theoretically and experimentally since the appearance of high intensity and
ultrashort pulse-lasers \cite{2ccc}. Up to the past decade, HHG phenomenon has
been extensively\ investigated in atomic and molecular gases \cite{hhg2}.
The intensity of the gaseous harmonics is relatively weak because of low gas
density, so that in the recent years became actual the nonperturbative study
of optical phenomena such as HHG in the condensed matter-solids. Currently,
the investigations of HHG and related nonlinear processes have been
successfully extended to bulk crystals \cite%
{att,imig1,imig2,imig3,sol4,sol5,sol6,sol7,sol10,sol11,sol12} and
two-dimensional (2D) nanostructures, such as graphene and graphene-like
materials \cite%
{H2,Mer,Mer1,H3,H4,H6,H7,H8,H9,H10,H11,H12,H12a,H13,H14,H15,H16,H17,H18},
monolayer transition metal dichalcogenides \cite{TMD}, \cite{TMD1},
hexagonal boron nitride \cite{BN}, topological insulators \cite{TI}, buckled
2D hexagonal nanostructures \cite{Mer2019}, and also solids \cite%
{corcumsolid}. Under some conditions, the process of HHG in bulk solids is
similar to atomic HHG \cite{sol7} and can be well described within a
semiclassical three-step model \cite{Corcum}, \cite{32b}. According to this
model the electron-hole is created, then accelerated, and in the last step
re-collided. At that, the substantial contribution to the HHG process gives
the interband current. The experiments \cite{sol4,sol5,sol6}, \cite%
{exp29,exp30,exp31} evidence this fact in solids. The experimental
verification of HHG phenomenon in a single layer transition metal
dichalcogenides \cite{sol11} and graphene \cite{H12} open up a wide way to
the higher harmonics radiation in 2D nanostructures. The appearance of a new
nano-optical materials-graphene-like nanostructures-with a very high carrier
mobility and extraordinary properties, such as graphene \cite{1}, \cite{2}
contributes using 2D nanostructures for nonlinear optical applications.

The experiments related to HHG process via THz pump wave pulses reported
about weak signals of harmonics \cite{H4} or lack \cite{exp41} of nonlinear
response in graphene. This is connected with the enormous fast relaxation of
electrons in graphene \cite{exp42} that prevents nonlinear optical processes
in the THz frequency region. So, the HHG experiment is considered with the
mid-IR light, where the generation of up to ninth harmonic in graphene has
been reported \cite{H12}. Theoretically, HHG has been investigated with even
mid-IR light in buckled graphene-like 2D nanostructures \cite{Mer2019},
where it has been shown that with the static electric field applied
perpendicular to the nanostructure sheet one can control the topology of the
bands and, as a result, one can generate even and odd high harmonics of
arbitrary polarization excited by a single-color infrared pump field. In 
\cite{Avet19}, the multicolor harmonic generation and wave-mixing nonlinear
processes in the efficient 2D nanostructures is studied. The presence of the
second laser field provides an additional flexibility for the implementation
of multiphoton excitation \cite{exp43}, \cite{exp44} expanding the spectrum
of possible combinations.

Among the mentioned materials, the bilayer graphene ($AB$-stacked) \cite{2}, 
\cite{1a}, \cite{1aa} possessing many interesting properties of monolayer
graphene \cite{9,99,22b,24b}, provides a richer band structure. The interlayer
coupling between the two graphene sheets changes the monolayer's Dirac cone,
inducing trigonal warping on the band dispersion and changing the topology
of the Fermi surface. This significantly enhances the rates of HHG \cite{H3}
in the THz region compared to monolayer graphene. Studies of the nonlinear
coherent response in $AB$-stacked bilayer graphene under the influence of
intense EM radiation also includes modification of quasi-energy spectrum,
the induction of valley polarized currents \cite{26b},\cite{27b}, as well as
second- and third-order nonlinear-optical effects \cite{28b,29b,30b,31b}.
The important advantage of bilayer graphene over the monolayer one is the
possibility to induce large tunable band gaps under the application of a
symmetry-lowering perpendicular electric field \cite{9}, \cite%
{10,16,HHGarxiv}.

In the present paper, we develop a nonlinear theory of the $AB$-stacked
bilayer graphene interaction with the coherent EM radiation, apart from the
advantages of the induced gap. For the second-order nonlinearity, one should
also take into account the trigonal warping of the bands. We consider a
multiphoton interaction in the nonperturbative regime. To exclude the
interband transitions, we have considered high Fermi energies, and an
external EM wave is taken to be from THz to the mid-IR domain of
frequencies. We show that there is an intense emission of harmonics at the
moderate intensities of the pump wave.

The paper is organized as follows. In Sec. II, the set of equations for a
single-particle density matrix is formulated and solved both analytically
and numerically in the multiphoton interaction regime. In Sec. III, we
consider the problem of harmonic generation at the multiphoton excitation of 
$AB$-stacked bilayer graphene at high Fermi energies in the low-frequency strong
coherent radiation field. Finally, conclusions are given in Sec. IV.

\section{Basic model}

In this paper we consider the multiphoton regime of harmonics generation in
a bilayer graphene by coherent radiation with frequency $\omega $ from THz
to the mid-IR and electric field amplitude $E_{0}$, at high Fermi energies
of graphene electrons. The wave-particle interaction at photon energies $%
\hbar \omega >\mathcal{E}_{L}$\ for intraband transitions can be
characterized by the dimensionless parameter $\chi =eE_{0}/(\omega \sqrt{%
m\hbar \omega })$. Here $\mathcal{E}_{L}=m\mathrm{v}_{3}^{2}/2\simeq 1$ $%
\mathrm{meV}$ -is the Lifshitz transition energy, $m=\gamma _{1}/(2\mathrm{v}_{F}^{2})$ is
the effective mass, $\mathrm{v}_{3}=\sqrt{3}a\gamma _{3}/(2\hbar )\approx 
\mathrm{v}_{F}/8$ is the effective velocity, $a\approx 0.246$ $\mathrm{nm}$
is the distance between the nearest $A$ sites, $\gamma
_{1}\simeq 0.39$ $\mathrm{eV}$, $\gamma _{3}=0.32$ $\mathrm{%
eV}$ is the interlayer hopping for the $AB$-stacked bilayer graphene; and $\mathrm{v}_{F}$ is the Fermi velocity
in a monolayer graphene. In considering case,
corresponding interaction parameter $\chi \gtrsim 1$. The intensity of the
wave can be estimated as\textrm{\ }%
\begin{equation}
I_{\chi }=\chi ^{2}\times 6\times 10^{10}\mathrm{Wcm}^{-2}(\hbar \omega /%
\mathrm{eV})^{3},  \label{23}
\end{equation}%
so the required intensity $I_{\chi }$\ for the nonlinear regime$\ $strongly
depends\textit{\ }on the parameter $\chi $ and photon energy $\hbar \omega $%
. Particularly, in the region from THz to the mid-IR photons with
wavelengths from $30$ $\mathrm{%
%TCIMACRO{\U{3bc} }%
%BeginExpansion
\mu
%EndExpansion
m}$ to $3$ $\mathrm{mm}$, the multiphoton interaction regime can be achieved
already at the values of the parameter $\chi \sim 0.5$\ corresponding to
pump wave intensities $I_{\chi }\sim 10$\textrm{\ k}$\mathrm{Wcm}^{-2}$. In
the opposite limit $\chi \ll 1$, the multiphoton effects are suppressed.

\begin{figure}[tbp]
\includegraphics[width=.7\textwidth]{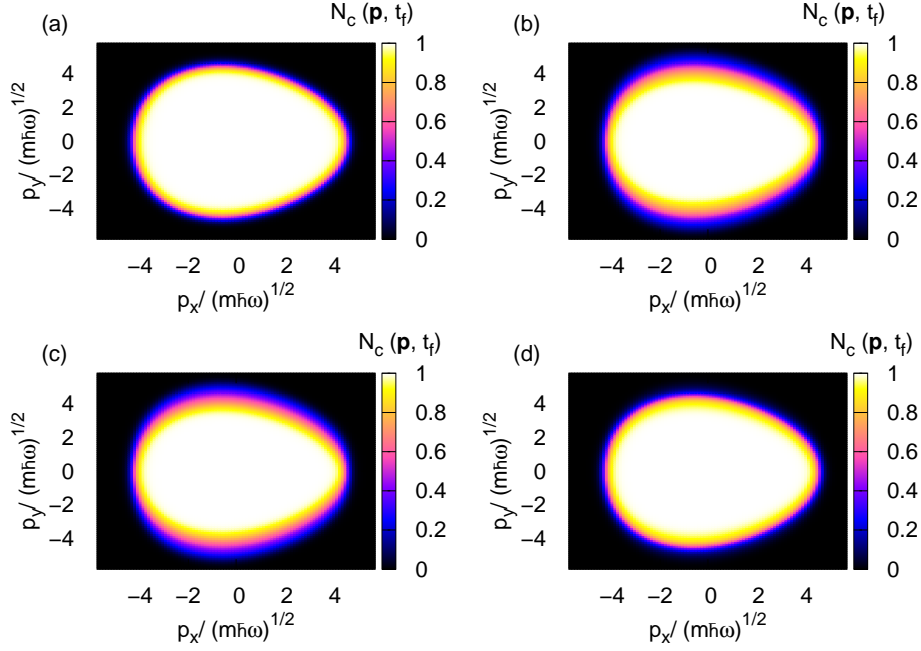}
\caption{(Color online) Particle distribution function $N_{c}(\mathbf{p}%
,t_{f})$\ (in arbitrary units) as a function of scaled dimensionless
momentum components is shown. The wave is assumed to be linearly polarized
along the $y$ axis. Multiphoton excitation with the trigonal warping effect
for the photon energy $\hbar \protect\omega $\ $=$\ $50$\ $\mathrm{meV}\simeq
50\mathcal{E}_{L} $, the temperature  $T/\hbar\protect\omega =0.5$ are demonstrated
at dimensionless intensity parameter $\protect\chi =1$ for valley $\protect%
\zeta =1$. (a)--(d) correspond to the density plot of the distribution function 
after the interaction at the instant $t_{f}=$0.25$\mathcal{T}$, 
 $t_{f}=$0.5$\mathcal{T}$,  $t_{f}=$0.75$\mathcal{T}$ and  $t_{f}=$$\mathcal{T}$, respectively.  }
\label{111}
\end{figure}

Let us consider the nonlinear microscopic theory of the multiphoton
interaction of bilayer graphene interaction with a coherent EM radiation. We
will discuss the ansatz of the evolutionary equation for a single-particle
density matrix. For the $AB$-stacked bilayer graphene, the low-energy
excitations $\left\vert \mathcal{E}\right\vert \ll \gamma _{1}\simeq 0.39$ $%
\mathrm{eV}$ in the vicinity of the Dirac points $\zeta K$ (valley quantum
number $\zeta =\pm 1$) can be described by an effective single-particle
Hamiltonian \cite{9,99,22b,24b}:

\begin{equation}
\widehat{H}_{\zeta }=\left( 
\begin{array}{cc}
0 & g_{\zeta }^{\ast }\left( \mathbf{p}\right)  \\ 
g_{\zeta }\left( \mathbf{p}\right)  & 0%
\end{array}%
\right) .  \label{1}
\end{equation}%
Here the operator $g_{\zeta }\left( \mathbf{p}\right) $\ is defined by the
relation 
\begin{equation}
g_{\zeta }\left( \mathbf{p}\right) =-\frac{1}{2m}\left( \zeta \widehat{p}%
_{x}+i\widehat{p}_{y}\right) ^{2}+\mathrm{v}_{3}\left( \zeta \widehat{p}%
_{x}-i\widehat{p}_{y}\right) ,  \label{2}
\end{equation}%
where $\mathbf{\hat{p}}=\left\{ \widehat{p}_{x},\widehat{p}_{y}\right\} $%
\textbf{\ }is the electron momentum operator. The first term in Eq. (\ref{2})
corresponds to a pair of parabolic bands $\mathcal{E}=\pm p^{2}/(2m)$, and
the second term coming from $\gamma _{3}$ causes trigonal warping in the
band dispersion. In the low-energy region $\hbar \omega <\mathcal{E}_{L}$
the two touching parabolas are transformed into the four separate
\textquotedblleft pockets\textquotedblright\ \cite{9}. The spin and the
valley quantum numbers are conserved. There is no degeneracy upon the valley
quantum number $\zeta $ for the issue considered. However, since there are
no intervalley transitions, the valley index $\zeta $ can be considered as a
characteristic parameter.

\begin{figure}[tbp]
\includegraphics[width=.45\textwidth]{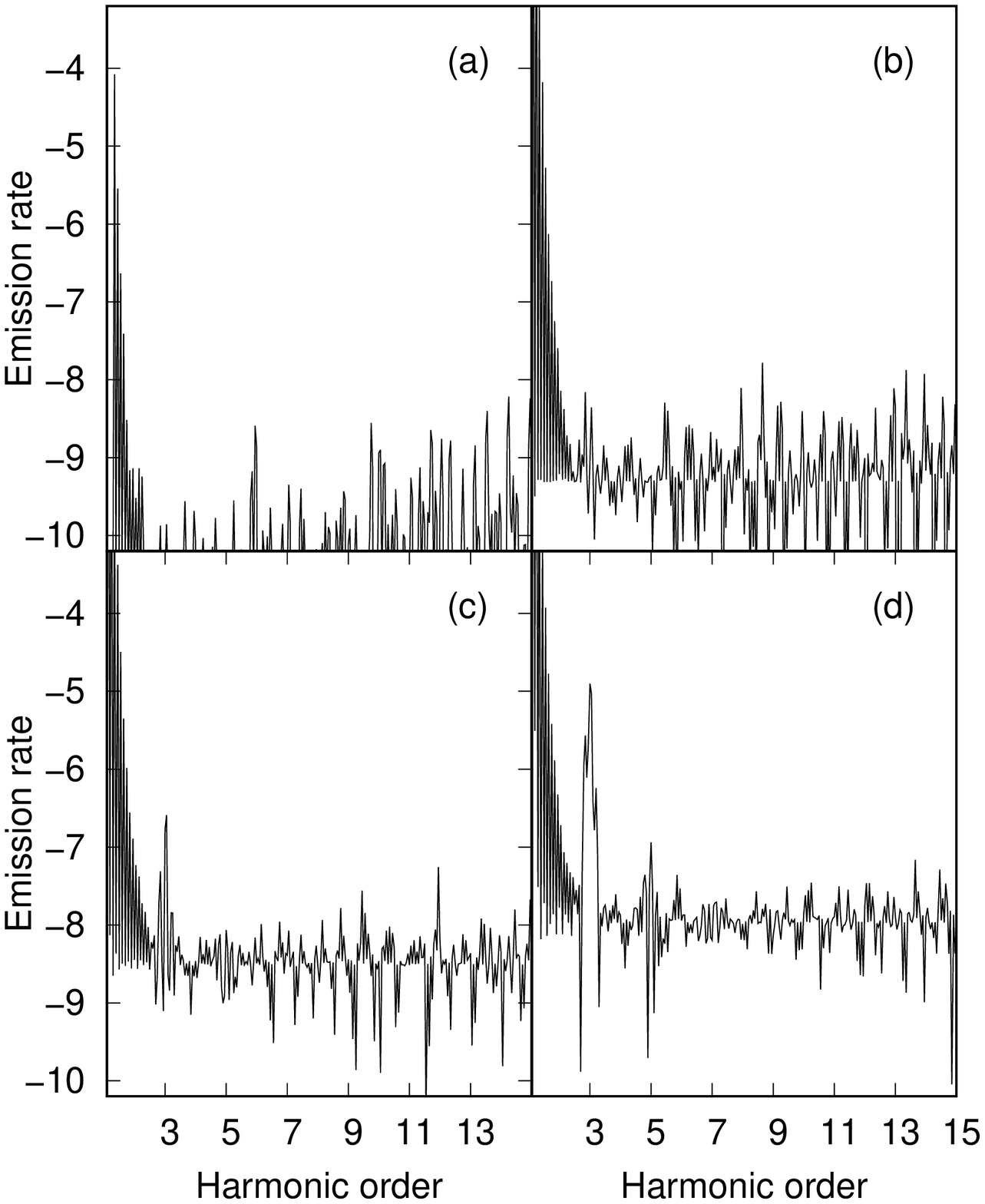}
\caption{ Harmonic emission rate in bilayer graphene at multiphoton
excitation via $log_{10}(n^{2}|J_{n}|^{2})$ (in arbitrary units), as a
function of the photon energy (in units of $\hbar \protect\omega $), is
shown for various intesities. The temperature is taken to be $T/\hbar 
\protect\omega =0.5$. The wave is assumed to be linearly polarized ($\protect%
\phi =0$) with frequency $\protect\omega =$\ $50$\ $\mathrm{meV/\hbar }$.
The results are for the intensity parameter (a) $\protect\chi =0.5$ , (b) $%
\protect\chi =1$ , (c) $\protect\chi =1.5$ , and (d) $\protect\chi =2$ . }
\label{222}
\end{figure}

The eigenstate functions of the effective Hamiltonian (\ref{1}) are the
spinors,%
\begin{equation}
\Psi _{\sigma }(\mathbf{r})=\frac{1}{\sqrt{S}}|\sigma ,\mathbf{p}\rangle e^{%
\frac{i}{\hbar }\mathbf{pr}},  \label{3}
\end{equation}%
where%
\begin{equation}
|\sigma ,\mathbf{p}\rangle =\frac{1}{\sqrt{2S}}\left( 
\begin{array}{c}
1 \\ 
\frac{1}{\mathcal{E}_{\sigma }}\Upsilon \left( \mathbf{p}\right)%
\end{array}%
\right) ,  \label{4}
\end{equation}%
with the eigenenergies:%
\begin{equation}
\mathcal{E}_{\sigma }\left( \mathbf{p}\right) =\sigma \sqrt{\left( \mathrm{v}%
_{3}p\right) ^{2}-\zeta \frac{\mathrm{v}_{3}p^{3}}{m}\cos 3\vartheta +\left( 
\frac{p^{2}}{2m}\right) ^{2}}.  \label{5}
\end{equation}%
Here, $\sigma $ is the band index: $\sigma =1$ and $\sigma =-1$ for
conduction and valence bands, respectively; $\vartheta $ is the angle $%
\vartheta =\arctan \left( p_{y}/p_{x}\right) $, $S$ is the quantization
area, and%
\begin{equation}
\Upsilon \left( \mathbf{p}\right) =-\frac{p^{2}}{2m}e^{i2\zeta \vartheta
}+\zeta \mathrm{v}_{3}pe^{-i\zeta \vartheta }.  \label{6}
\end{equation}

We will investigate the case when the bilayer graphene interacts with a
plane quasimonochromatic EM radiation of carrier frequency $\omega $ and
slowly varying envelope. To exclude the effect of the magnetic field, the
wave is taken propagating in the perpendicular direction to the graphene
sheets ($XY$). To investigate the dependence on the wave polarization, the
pump wave is taken to be elliptically polarized with $\mathbf{E}\left(
t\right) $: 
\begin{equation}
\mathbf{E}\left( t\right) =f\left( t\right) E_{0}\left( \widehat{\mathbf{x}}%
\sin \phi \cos \omega t+\widehat{\mathbf{y}}\cos \phi \sin \omega t\right) .
\label{7}
\end{equation}%
The slowly varying envelope of the wave describes by the $sin$-squared
envelope function:%
\begin{equation}
f\left( t\right) =\left\{ 
\begin{array}{cc}
\sin ^{2}\left( \pi t/\mathcal{T}\right) , & 0\leq t\leq \mathcal{T} \\ 
0, & t<0,t>\mathcal{T}%
\end{array}%
\right. ,  \label{8}
\end{equation}%
where $\mathcal{T}$ characterizes the pulse duration and is taken to be
ten wave cycles: $\mathcal{T}=10\mathcal{T}_{0}$; $\phi $ is the pump
wave polarization parameter, $\mathcal{T}_{0}=2\pi /\omega $ is the wave
period.

To follow the second quantization formalism, expanding the fermionic field
operators on the basis of the states $\Psi _{\sigma }(\mathbf{r})$ (\ref{4}%
), 
\begin{equation}
\widehat{\Psi }(\mathbf{r},t)=\sum\limits_{\mathbf{p,}\sigma }\widehat{a}_{%
\mathbf{p},\sigma }(t)\Psi _{\sigma }(\mathbf{r}),  \label{9}
\end{equation}%
where $\widehat{a}_{\mathbf{p},\sigma }^{+}(t)$ ($\widehat{a}_{\mathbf{p}%
,\sigma }(t)$) is the creation (annihilation) operator of an electron with
the momentum $\mathbf{p}$\textbf{,} which satisfies the usual fermionic
anticommutation rules at equal times. Under the influence of a uniform
time-dependent electric field $E(t)$ the single-particle Hamiltonian can be
expressed in the form:%
\begin{equation}
\widehat{H}=\widehat{H}_{\zeta }+\left( 
\begin{array}{cc}
e\mathbf{rE}\left( t\right)  & 0 \\ 
0 & e\mathbf{rE}\left( t\right) 
\end{array}%
\right) ,  \label{12}
\end{equation}%
where the light-matter interaction Hamiltonian is taken in the length gauge 
\cite{Corcum}, \cite{32b} that provides proper inclusion of the inter- and
intraband transitions \cite{Avet19}. Taking into account an expansion (\ref%
{9}), the second quantized total Hamiltonian can be expressed as: 
\begin{equation}
\widehat{H}=\sum\limits_{\sigma ,\mathbf{p}}\mathcal{E}_{\sigma }\left( 
\mathbf{p}\right) \widehat{a}_{\sigma \mathbf{p}}^{+}\widehat{a}_{\sigma 
\mathbf{p}}+\widehat{H}_{\mathrm{s}},  \label{13}
\end{equation}%
where the light--matter interaction part is given in terms of the field $%
\mathbf{E}\left( t\right) $ as follow:%
\begin{equation*}
\widehat{H}_{\mathrm{s}}=ie\sum\limits_{\mathbf{p,p}^{\prime },\sigma
}\delta _{\mathbf{p}^{\prime }\mathbf{p}}\partial _{\mathbf{p}^{\prime }}%
\mathbf{E}\left( t\right) \widehat{a}_{\mathbf{p},\sigma }^{\dagger }%
\widehat{a}_{\mathbf{p}^{\prime },\sigma ^{\prime }}\ 
\end{equation*}%
\begin{equation}
+\sum\limits_{\mathbf{p},\sigma }\mathbf{E}\left( t\right) \left( \mathbf{D}%
_{\mathrm{t}}\left( \sigma ,\mathbf{p}\right) \widehat{a}_{\mathbf{p},\sigma
}^{+}\widehat{a}_{\mathbf{p},-\sigma }+\mathbf{D}_{\mathrm{m}}\left( \sigma ,%
\mathbf{p}\right) \widehat{a}_{\mathbf{p},\sigma }^{+}\widehat{a}_{\mathbf{p}%
,\sigma }\right) .  \label{14}
\end{equation}%
Here the quantity $\mathbf{D}_{\mathrm{t}}\left( \sigma ,\mathbf{p}\right) $%
, 
\begin{equation}
\mathbf{D}_{\mathrm{t}}\left( \sigma ,\mathbf{p}\right) =\hbar e\langle
\sigma ,\mathbf{p}|i\partial _{\mathbf{p}}|-\sigma ,\mathbf{p}\rangle 
\label{15}
\end{equation}%
is the transition dipole moment and $\mathbf{D}_{\mathrm{m}}\left( \sigma ,%
\mathbf{p}\right) $ 
\begin{equation}
\mathbf{D}_{\mathrm{m}}\left( \sigma ,\mathbf{p}\right) =\hbar e\langle
\sigma ,\mathbf{p}|i\partial _{\mathbf{p}}|\sigma ,\mathbf{p}\rangle 
\label{16}
\end{equation}%
is the Berry connection or the mean dipole moment. Note that the matrix
elements (\ref{15}), (\ref{16}) are actually gauge-dependent \cite{Mer2019}.

We will represent the multiphoton interaction of a bilayer graphene with a
coherent EM wave field by the Liouville--von Neumann equation for a
single-particle density matrix%
\begin{equation}
\rho _{\alpha ,\beta }(\mathbf{p},t)=\langle \widehat{a}_{\mathbf{p},\beta
}^{+}\left( t\right) \widehat{a}_{\mathbf{p},\alpha }\left( t\right) \rangle
.  \label{17}
\end{equation}%
Here $\widehat{a}_{\mathbf{p},\alpha }\left( t\right) $ obeys the Heisenberg
equation 
\begin{equation}
i\hbar \frac{\partial \widehat{a}_{\mathbf{p},\alpha }\left( t\right) }{%
\partial t}=\left[ \widehat{a}_{\mathbf{p},\alpha }\left( t\right) ,\widehat{%
H}\right] .  \label{18}
\end{equation}%
The evolutionary equation for the single-particle density matrix will write
in the form, using Eqs. (\ref{13})-(\ref{18}):  
\begin{equation*}
i\hbar \frac{\partial \rho _{\alpha ,\beta }(\mathbf{p},t)}{\partial t}%
-i\hbar e\mathbf{E}\left( t\right) \frac{\partial \rho _{\alpha ,\beta }(%
\mathbf{p},t)}{\partial \mathbf{p}}=
\end{equation*}%
\begin{equation*}
\left. \left( \mathcal{E}_{\alpha }\left( \mathbf{p}\right) -\mathcal{E}%
_{\beta }\left( \mathbf{p}\right) -i\hbar \Gamma \left( 1-\delta _{\alpha
\beta }\right) \right) \rho _{\alpha ,\beta }(\mathbf{p},t)\right] 
\end{equation*}%
\begin{equation*}
+\mathbf{E}\left( t\right) \left( \mathbf{D}_{\mathrm{m}}\left( \alpha ,%
\mathbf{p}\right) -\mathbf{D}_{\mathrm{m}}\left( \beta ,\mathbf{p}\right)
\right) \rho _{\alpha ,\beta }(\mathbf{p},t)
\end{equation*}%
\begin{equation}
+\mathbf{E}\left( t\right) \left[ \mathbf{D}_{\mathrm{t}}\left( \alpha ,%
\mathbf{p}\right) \rho _{-\alpha ,\beta }(\mathbf{p},t)-\mathbf{D}_{\mathrm{t%
}}\left( -\beta ,\mathbf{p}\right) \rho _{\alpha ,-\beta }(\mathbf{p},t)%
\right] .  \label{19}
\end{equation}%
Here $\Gamma $ is the damping rate. Note that due to the homogeneity of the
problem we only need the $\mathbf{p}$-diagonal elements of the density
matrix. We will also incorporate relaxation processes into Liouville--von
Neumann equation with inhomogeneous phenomenological damping term, since
homogeneous relaxation processes are slow compared with inhomogeneous one.

In Eq. (\ref{19}) the diagonal elements represent particle distribution
functions for conduction $N_{c}(\mathbf{p},t)=\rho _{1,1}(\mathbf{p},t)$ and
valence $N_{\mathrm{v}}(\mathbf{p},t)=\rho _{-1,-1}(\mathbf{p},t)$ bands,
and the nondiagonal elements are interband polarization $\rho _{1,-1}(%
\mathbf{p},t)=P(\mathbf{p},t)$ and its complex conjugate $\rho _{-1,1}(%
\mathbf{p},t)=P^{\ast }(\mathbf{p},t)$. As an initial state ($t=0$) we
assume an ideal Fermi gas in equilibrium with particle distribution function%
\begin{equation}
\rho _{\sigma ,\sigma ^{\prime }}(\mathbf{p},0)=\frac{1}{1+e^{\left[ 
\mathcal{E}_{\sigma }\left( \mathbf{p}\right) -\mu \right] /T}}\delta
_{\sigma ,\sigma ^{\prime }},  \label{ddd}
\end{equation}%
where $\mu $ $\left( \mu \simeq \varepsilon _{F}\right) $ and $T$ are the
chemical potential and the temperature, respectively, in energy units.

It is clear that a strong field will induce multiphoton transitions from the
valence to the conduction band. Thus, because of Pauli blocking and low
frequency of driving wave the interband transitions can take place via
multiphoton channels: $n_{0}\hbar \omega >2\varepsilon _{F}$. We will
consider the case when $2\varepsilon _{F}\gg \hbar \omega $ and consequently 
$n_{0}\gg 1$. This is a quasiclassical regime when the wave-particle
interaction can be characterized by the work of the wave electric field
during the wave period $eE_{0}\mathrm{v}_{x}/\omega $. To neglect the
interband transitions the latter should be smaller than $2\varepsilon _{F}:$ 
$eE_{0}\mathrm{v}_{x}/\omega <2\varepsilon _{F}$. This means that the
wavefield can not provide sufficient energy for the creation of an
electron-hole pair.

Thus, in particular case of a pump EM pulse with THz up to mid-IR frequencies
and for high Fermi energies the interband transitions are excluded, one can
simplify the closed set of the differential equations Eq. (\ref{19}). In
addition, we will assume that the system relaxes at a rate $\gamma $ to the
equilibrium $N_{c}^{(0)}(\mathbf{p},t)$ distribution. Thus, from (\ref{19})
we obtain the Boltzmann equation for the electrons distribution function $%
N_{c}(\mathbf{p},t)$:%
\begin{equation}
i\hbar \frac{\partial N_{c}(\mathbf{p},t)}{\partial t}-i\hbar e\mathbf{E}%
\left( t\right) \frac{\partial N_{c}(\mathbf{p},t)}{\partial \mathbf{p}}%
=-\gamma \left( N_{c}(\mathbf{p},t)-N_{c}^{(0)}(\mathbf{p},t)\right) .
\label{20}
\end{equation}%
We construct $N_{c}^{(0)}(\mathbf{p},t)$ from the filling of electron states
according to the Fermi--Dirac-distribution Eq. (\ref{ddd}):%
\begin{equation}
N_{c}^{(0)}(\mathbf{p},t)=\frac{1}{1+e^{\left[ \mathcal{E}_{1}\left( \mathbf{%
p}\right) -\mu \right] /T}}.  \label{21}
\end{equation}%
Note that this relaxation approximation provides an accurate description for
the optical field components oscillating at frequencies $\omega \gg \gamma $
(in accordance with the relaxation rate $\Gamma \ll \mathcal{T}_{0}^{-1}$).

The Eq (\ref{20}) can be solved analytically by the method of
characteristics, then the solution has the form:%
\begin{equation}
N_{c}(\mathbf{p},t)=\gamma \int d\tau e^{-\gamma \left( t-\tau \right)
}\left( N_{c}^{(0)}(\mathbf{p+p}_{E}\left( t,\tau \right) \right),
\label{22}
\end{equation}%
where\ 
\begin{equation}
\mathbf{p}_{E}\left( t,\tau \right) =-e\int_{\tau }^{t}\mathbf{E}\left(
t^{"}\right) dt^{"}  \label{24}
\end{equation}%
is the momentum given by the wave field.

\begin{figure}[tbp]
\includegraphics[width=.45\textwidth]{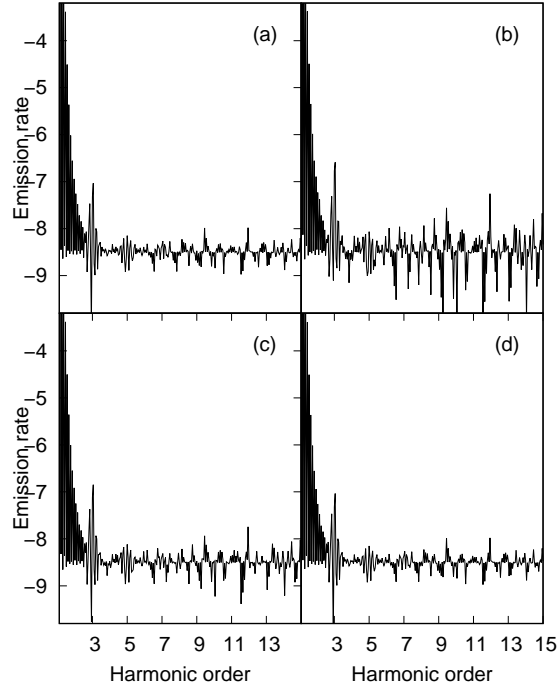}
\caption{High harmonic spectra for bilayer graphene at multiphoton
excitation is shown for various wave frequencies in logarithmic scale. The
temperature is taken to be $T/\hbar \protect\omega =0.5$.
The wave is assumed to be linearly polarized ($\protect\phi =0$) with the
intensity $\protect\chi =1.5$. The results are for frequency (a) $\protect%
\omega =40$\ $\mathrm{meV/\hbar }$, (b) $\protect\omega =50$\ $\mathrm{%
meV/\hbar }$, (c) $\protect\omega =60$\ $\mathrm{meV/\hbar }$, and (d) $%
\protect\omega =70$\ $\mathrm{meV/\hbar }$. }
\label{333}
\end{figure}

For the coherent part of the radiation spectrum, one needs the mean value of
the current density operator,%
\begin{equation}
j_{\zeta }=-2e\left\langle \widehat{\Psi }(\mathbf{r},t)\left\vert \widehat{%
\mathbf{v}}_{\zeta }\right\vert \widehat{\Psi }(\mathbf{r},t)\right\rangle ,
\label{50}
\end{equation}%
where $\widehat{\mathbf{v}}_{\zeta }=\partial \widehat{H}/\partial \widehat{%
\mathbf{p}}$ is the velocity operator and we have taken into account the
spin degeneracy factor $2$. For the effective $2\times 2$ Hamiltonian (\ref%
{1}) the velocity operator in components reads:%
\begin{equation}
\widehat{\mathrm{v}}_{\zeta x}=\zeta \left( 
\begin{array}{cc}
0 & -\frac{1}{m}\left( \zeta \widehat{p}_{x}-i\widehat{p}_{y}\right) +%
\mathrm{v}_{3} \\ 
-\frac{1}{m}\left( \zeta \widehat{p}_{x}+i\widehat{p}_{y}\right) +\mathrm{v}%
_{3} & 0%
\end{array}%
\right) ,  \label{51}
\end{equation}%
\begin{equation}
\widehat{\mathrm{v}}_{\zeta y}=i\left( 
\begin{array}{cc}
0 & \frac{1}{m}\left( \zeta \widehat{p}_{x}-i\widehat{p}_{y}\right) +\mathrm{%
v}_{3} \\ 
-\frac{1}{m}\left( \zeta \widehat{p}_{x}+i\widehat{p}_{y}\right) -\mathrm{v}%
_{3} & 0%
\end{array}%
\right) .  \label{52}
\end{equation}%
Using the Eqs. (\ref{50})--(\ref{52}) and (\ref{17}), the expectation value
of the current for the valley $\zeta $ can be written in the form:%
\begin{equation}
\mathbf{j}_{\zeta }\left( t\right) =-\frac{e}{2\pi ^{2}\hbar ^{2}}\int d%
\mathbf{pV}\left( \mathbf{p}\right) N_{c}(\mathbf{p},t)  \label{53}
\end{equation}%
where 
\begin{equation}
\mathbf{V}\left( \mathbf{p}\right) =\frac{\mathrm{v}_{3}\mathbf{p}-3\zeta 
\frac{\mathrm{v}_{3}p}{2m}\mathbf{p}\cos 3\vartheta +3\zeta \frac{\mathrm{v}%
_{3}p^{3}}{2m}\sin 3\vartheta \frac{\partial \vartheta }{\partial \mathbf{p}}%
+2\frac{\mathbf{p}^{3}}{\left( 2m\right) ^{2}}}{\mathcal{E}_{1}\left( 
\mathbf{p}\right) }  \label{Vp}
\end{equation}%
is the intraband velocity. As is seen from Eq. (\ref{53}), in considering
case the surface current provides a source for the generation of harmonic
radiation. It is the intraband current $\sim N_{c}(\mathbf{p},t)$. The
intraband high harmonics are generated as a result of the independent motion
of carriers in their respective bands. The interband high harmonics (given
by the term $N_{\mathrm{v}}(\mathbf{p},t)$) which are generated as a result
of the recombination of accelerated electron-hole pairs \cite{HHGarxiv}, are
unessential in the considering case.

Let us now consider a particular case of coherent interaction of a bilayer
graphene with a pump EM wave in the ultrafast excitation regime, which is
correct only for the times $t<\tau _{\min }$, where $\tau _{\min }$\ is the
minimum of all relaxation times. For the excitations of energies $\mathcal{E}%
\ll \gamma _{1}=0.39$\ $\mathrm{eV}$, the dominant mechanism for relaxation
will be electron-phonon coupling via longitudinal acoustic phonons \cite%
{Hwang}, \cite{Viljas}. For the low-temperature limit, if $T\ll 2\left(
c_{ph}/\mathrm{v}_{F}\right) \sqrt{\mathcal{E}\gamma _{1}}$, where $%
c_{ph}\simeq 2\times 10^{6}$\ $\mathrm{cm/s}$\ is the velocity of
longitudinal acoustic phonons, the relaxation time for the energy level $%
\mathcal{E}$\ can be estimated as \cite{Viljas}:%
\begin{equation}
\tau \left( \mathcal{E}\right) \simeq \left( \frac{\pi D^{2}T^{2}}{8\rho
_{m}\hbar ^{3}c_{ph}^{3}\mathrm{v}_{F}}\sqrt{\frac{\gamma _{1}}{\mathcal{E}}}%
\right) ^{-1}.  \label{form1}
\end{equation}%
For the high-temperature limit $T\gg 2\left( c_{ph}/v_{F}\right) \sqrt{%
\mathcal{E}\gamma _{1}}$ ($\mathcal{E}\simeq \mu =10\hbar \omega $, $T$ is
the room temperature) we can use the relation: 
\begin{equation}
\tau \simeq \left( \frac{D^{2}T\gamma _{1}}{4\rho _{m}\hbar ^{3}c_{ph}^{2}%
\mathrm{v}_{F}^{2}}\right) ^{-1}.  \label{form}
\end{equation}%
Here $D\simeq 20$\ $\mathrm{eV}$\ is the electron-phonon coupling constant,
and $\rho _{m}\simeq 15\times 10^{-8}$\ $\mathrm{g/cm}^{2}$\ is the mass
density of a bilayer graphene. For $\hbar \omega \simeq 0.05$\ $\mathrm{eV}$
and $\mathcal{E}\simeq \mu =10\hbar \omega $ at the temperatures $T=0.025$\ $%
\mathrm{eV}$ ($T\gg 2\left( c_{ph}/\mathrm{v}_{F}\right) \sqrt{\mathcal{E}%
\gamma _{1}}$), from Eq. (\ref{form}) we obtain $\tau \simeq 1.1$\ $\mathrm{%
ps}$. Thus, in this energy range one can coherently manipulate with the
multiphoton transitions in a bilayer graphene on time scales $t\precsim 1$\ $%
\mathrm{ps}$ neglecting the particle-particle collisions.

\section{HHG at the electrons intraband multiphoton excitations}

In this section, we will investigate the nonlinear response of a bilayer
graphene in the process of higher harmonics generation under the influence
of a pump EM wave with the frequencies in the domain from THz to mid-IR
ones: $\omega =0.04\div 0.07$ $\mathrm{eV/}\hbar $.

Taking into account Eqs. (\ref{53}) and (\ref{22}), the total intraband
current can be represented as:%
\begin{equation}
\mathbf{j}_{\varsigma }\left( t\right) =-\frac{\gamma e}{2\pi ^{2}\hbar ^{2}}%
\int_{-\infty }^{t}d\tau e^{-\gamma \left( t-\tau \right) }\int d\mathbf{pV}%
\left( \mathbf{p}\right) \left( N_{c}^{(0)}(\mathbf{p+p}_{E}\left( t,\tau
\right) \right) .  \label{54}
\end{equation}%
There is no degeneracy upon the valley quantum number $\zeta $, so the total
current is obtained by a summation over $\zeta $: 
\begin{equation}
j_{x}=j_{1,x}+j_{-1,x};  \label{55}
\end{equation}%
\begin{equation}
j_{y}=j_{1,y}+j_{-1,y}.  \label{56}
\end{equation}%
From Eq. (\ref{54}) we see that 
\begin{equation}
\frac{j_{x,y}}{j_{0}}=J_{x,y}\left( \omega t,\chi ,\frac{\gamma }{\omega },%
\frac{\mathcal{E}_{L}}{\hbar \omega },\frac{\mu }{\hbar \omega },\frac{T}{%
\hbar \omega }\right) ,  \label{57}
\end{equation}%
where 
\begin{equation}
j_{0}=\frac{e\omega }{\pi ^{2}}\sqrt{\frac{m\omega }{\hbar }},  \label{58}
\end{equation}%
and $J_{x}$ and $J_{y}$ are the dimensionless periodic (for monochromatic
wave) functions which parametrically depend on the interaction parameter $%
\chi $, scaled Lifshitz energy, and macroscopic parameters of the system.
Thus, having solutions of \ Eq. (\ref{20}) and making integration in Eq. (%
\ref{54}), one can calculate the harmonic radiation spectra with the help of
a Fourier transform of the function $J_{x,y}(t)$. The emission rate of the $n
$th harmonic is proportional to $n^{2}|j_{n}|^{2}$, where $%
|j_{n}|^{2}=|j_{xn}|^{2}+|j_{yn}|^{2}$, with $j_{xn}$ and $j_{yn}$ being the 
$n$th Fourier components of the field-induced total current. To find $j_{n}$%
, the fast Fourier transform algorithm has been used. We have used the
normalized current density (\ref{57}) for the plots.

For the following investigations, we made a change of variables and
transform the equations with partial derivatives into ordinary ones. The new
variables are $t$ and $\widetilde{\mathbf{p}}=\mathbf{p}-\mathbf{p}%
_{E}\left( t\right) .$After the simple transformations, the integration of
equation (\ref{20}) is performed on a homogeneous grid of $10^{4}$ ($%
\widetilde{p}_{x},\widetilde{p}_{y}$)-points. For the maximal momentum, we
take $\widetilde{p}_{\max }/\sqrt{m\hbar \omega }=5.8$. The time integration
is performed numerically with the standard fourth-order Runge-Kutta
algorithm. For numerical analysis of HHG rates in bilayer graphene, we
assume high Fermi energy $\varepsilon _{F}\simeq \mu =10\hbar \omega $ ($%
\varepsilon _{F}\gg \hbar \omega $). The damping rate $\gamma /\omega =0.5$
will be assumed in all plots below.

In Fig. 1-5 the temperature is taken $T=0.025$\ $\mathrm{eV}$, and the pump
wave is assumed to be linearly polarized along with the $Y$ axis. In Fig. 1,
the wave pulse duration for $\omega =10\mathcal{E}_{L}/\hbar =50$ $\mathrm{%
meV}$ is $\mathcal{T}=10\mathcal{T}_{0}\simeq 0.82$ $\mathrm{ps}$.
Photoexcitations of the Fermi-Dirac sea are presented in Fig.1, where the
 density plot of the particle
distribution function $N_{c}(\mathbf{p},t_{f})$\ is shown as a function of
scaled dimensionless momentum components after the interaction at the different instances of the pump wave 
pulse duration. The picture partially
changes with the on and off the pulse of the wave. This figure shows the
electrons' intraband transition only. The nonlinear trigonal warping effect
describing the deviation of the excited iso-energy contours from circles is
seen clearly.

In Fig. 2 the high harmonic spectra for bilayer graphene at multiphoton
excitation is shown for various wave intensities. As it is seen, the
contribution of higher-order harmonics in emission rate is more significant
with the intensity increase. The analysis also shows the linear dependence
of the harmonics number cutoff on the amplitude of a pump electric field $%
n_{coff}\sim \chi $. Note that in the considering case of an intrinsic
gapless bilayer, the system possesses in-plane inversion symmetry, and at
the normal incidence of the pump wave on the bilayer graphene only odd
harmonics are generated \cite{Mer}.

\begin{figure}[tbp]
\includegraphics[width=.45\textwidth]{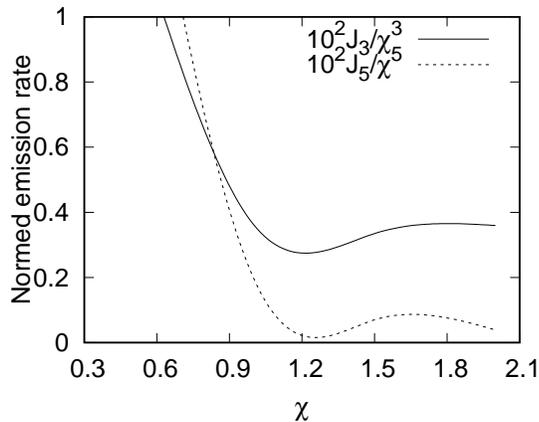}
\caption{Third ($J_{3}/\protect\chi ^{3}$) and fivth ($J_{5}/\protect\chi %
^{5}$) harmonics scaled emission rate (in arbitrary units) for bilayer
graphene versus $\protect\chi $. The temperature is taken to be $T/\hbar 
\protect\omega =0.5$. The wave is assumed to be linearly polarized along the 
$y$ axis with the frequency $\protect\omega =$\ $50$\ $\mathrm{meV/\hbar }$. 
}
\label{444}
\end{figure}

For clarification of the harmonics generation regime, we examine the
emission rate of the higher harmonics versus pump wave strength $\chi $ at the
same wave intensity $I_{\chi }\simeq 1.7\times 10^{7}\mathrm{\ Wcm}^{-2}$
for various frequencies, which is shown in Fig. 3. Third  ($J_{3}/\protect\chi ^{3}$) and fifth  ($J_{5}/\protect\chi ^{5}$)
 harmonics scaled emission rate for bilayer graphene versus $\protect\chi $ is demonstrated in Fig. 4. As is seen from this
figure, up to the field strengths $\chi <1$ we almost have power-law
 for the emission rate in accordance with the perturbation theory. For
large $\chi $ we have a strong deviation from power law for the emission
rate of high harmonics. The temperature dependence is
demonstrated in Fig. 5. This investigation shows that in intrinsic bilayer
graphene \cite{H3} the harmonics are suppressed at high temperatures. As
show the plots of Fig. 2-5 at high Fermi energies the HHG rates are larger
compare with the intrinsic bilayer graphene with taking into account the
interband transitions with intraband ones.

\begin{figure}[tbp]
\includegraphics[width=.45\textwidth]{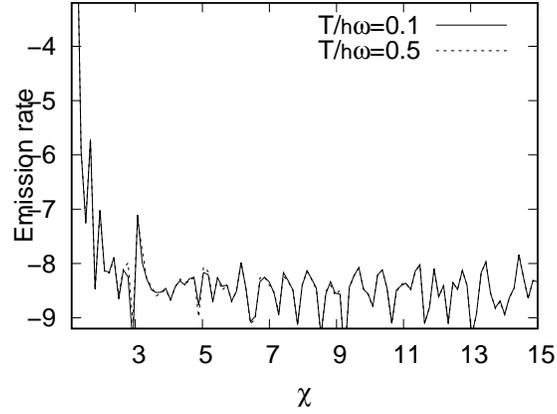}
\caption{ High harmonic spectra for bilayer graphene at
multiphoton excitation for a linearly polarized wave ($\protect\phi =0$) is
shown for temperatures $T/\hbar \protect\omega =0.1 $ and $T/\hbar \protect%
\omega =0.5$. The wave intensity $\protect\chi =1.5$ and frequency $\protect%
\omega =50$\ $\mathrm{meV/\hbar }$. }
\label{555}
\end{figure}

\begin{figure}[tbp]
\includegraphics[width=.45\textwidth]{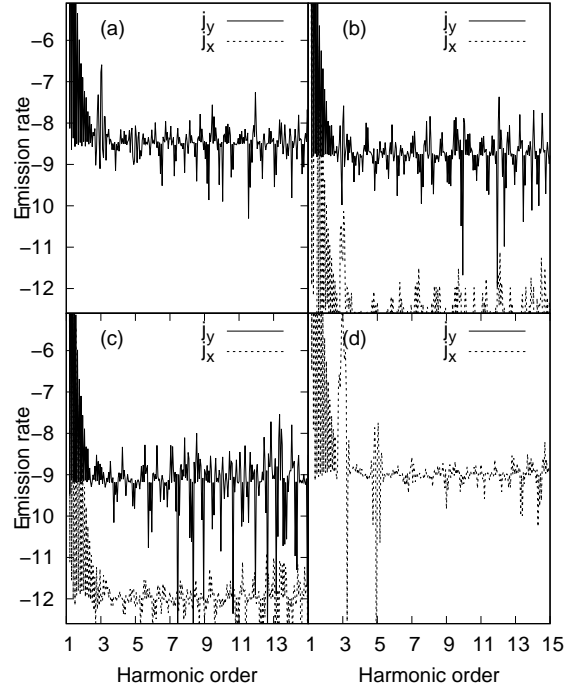}
\caption{High harmonic spectra in logarithmic scale for a elliptically
polarized wave is shown at temperature $T/\hbar \protect\omega =0.5$ at the
wave intensity $\protect\chi =1.5$ and frequency $\protect\omega =50$\ $%
\mathrm{meV/\hbar }$. The results are for (a) $\protect\phi =0$%
, (b) $\protect\phi =\protect\pi /6$, (c) $\protect\phi =\protect\pi /4$,
and (d) $\protect\phi =\protect\pi /2$, respectively. }
\label{666}
\end{figure}

Finally in Fig. 6, we show the dependence of HHG on the polarization of the
pump wave. The results are for linearly polarized wave along the $x$ ($\phi
=\pi /2$) and $y$ ($\phi =0$) axes, for circular polarization ($\phi =\pi /4$%
) and for elliptic polarization ($\phi =\pi /6$). As is seen, orienting the
linearly polarized pump wave along with these axes results\textit{\ }in
different harmonics spectra. This is because we have strongly anisotropic
excitation near the Dirac points. For elliptic and circular polarizations
the rates for the middle harmonics increase, while high order harmonics are
suppressed.

As was mentioned in \cite{H3}, the current amplitude (\ref{58}) compare to
intrinsic graphene $j_{0}$ \cite{Mer}, \cite{Mer1}, for bilayer graphene is
larger by a factor $\left( \gamma _{1}/2\hbar \omega \right) ^{1/2}$.
Besides, the cutoff harmonic is larger than in the case of monolayer
graphene \cite{Mer}, which is a result of strong nonlinearity caused by
trigonal warping. Hence, for considered setups $\hbar \omega \ll \gamma _{1}$
the harmonics' radiation intensity is at least one order of magnitude larger
than in the monolayer graphene.

Here we will estimate the conversion efficiency for harmonics $\eta
_{n}=I_{n}/I$ as in \cite{H3} 
\begin{equation*}
\eta _{n}\sim 10^{-3}\chi ^{-2}\left( d/\lambda \right) ^{2}\left(
nJ_{n}\right) ^{2},
\end{equation*}%
where $\lambda =2\pi c/\omega $ and $d$ is the characteristic size of the
bilayer graphene sheet. For the setup of Fig. 2 at the intensity parameter $%
\chi \simeq 1$, depending on the ratio, even for the $d\sim \lambda $, one
can achieve conversion efficiencies $\eta _{n}\sim 10^{-2}$ for up to the
ninth harmonic. Note that these quite large conversion efficiencies are
obtained for high order harmonics in the case of the single bilayer graphene
sheet. For the low order harmonics, we have $\eta _{3}\sim 10^{-7}$, and for
the experimental realization one can use multilayer $N_{l}\gg 1$ sheet \cite%
{cascadlaser} up to experimentally achievable values $N_{l}\sim 1$
monolayers \cite{cascad} with the thickness $\sim 20$ $\mathrm{nm}$. Since
film thickness is much smaller than the considered wavelengths, the
harmonics' signal from all layers will sum up constructively. Thus, for the
average conversion efficiencies we will have the same magnitude $\sim 0.02$.
In an experiment, one can use many layers, which are comparable to what one
expects to achieve with resonant two-level systems in nonlinear optics \cite%
{nonlinopt}.

\section{Conclusion}

On the base of the microscopic quantum theory of nonlinear interaction of a
bilayer graphene with a coherent EM radiation at high Fermi energies of
electrons towards the high harmonics generation has been investigated. The
differential equations for the single-particle density matrix is solved both
analytically and numerically in the vicinity of $\zeta K$ points in the
Brillouin zone. We have considered the practically interesting regimes of
multiphoton excitation of the Fermi-Dirac sea for effective generation of
harmonics via the pump wave pulses from THz to the mid-IR domain of
frequencies. The considered domains of frequencies and high Fermi energies
exclude the valence band excitations and interband transitions. The cutoff
of harmonics in these regimes increases with the pump wave intensity
enhancement and harmonics emission processes become robust against the
temperature increase. As a result of the strong nonlinearity caused by the
trigonal warping, the current amplitude for bilayer graphene is at one order
of the magnitude larger than in the intrinsic graphene. Moreover, it has
been shown strict growing of HHG rates in considering case compare to the
case of HHG in bilayer graphene with the intra- and interband multiphoton
transitions \cite{HHGarxiv}. The obtained results show that bilayer graphene
may serve as an effective medium for generation of higher harmonics at room
temperatures by the pump waves from THz to mid-IR frequencies and intensitis 
$\sim $ $10$\textrm{\ k}$\mathrm{Wcm}^{-2}$. In this context, the
requirement of high intensity in the THz regime does not preclude the use of
standard THz lasers, which are available \cite{laser}. Moreover, if in the
THz domain of pump wave frequencies reported about weak signals at HHG \cite%
{H4} concerning the generation of high harmonics up to the mid-IR range,
note that this has been demonstrated in the paper \cite{HHGarxiv} where
Quantum Cascade lasers \cite{laser1} are readily available and can provide
higher powers.

\begin{acknowledgments}
The authors are deeply grateful to prof. H. K. Avetissian for permanent discussions and valuable recommendations.
This work was supported by the RA MES Science Committee.
\end{acknowledgments}


\begin{thebibliography}{99}
\bibitem{1Corkum} P. B. Corkum and F. Krausz , \textquotedblright Attosecond
science\textquotedblright , \textit{Nature Physics} \textbf{3} 381-387
(2007), https://doi.org/10.1038/nphys620.

\bibitem{hhg1} T. Brabec and F. Krausz, \textquotedblright Intense few-cycle
laser fields: Frontiers of nonlinear optics\textquotedblright , \textit{Rev.
Mod. Phys.} \textbf{72}, 545--591 (2000),
https://doi.org/10.1103/RevModPhys.72.545.

\bibitem{Abook} H. K. Avetissian, \textquotedblright Relativistic Nonlinear
Electrodynamics\textquotedblright , The QED vacuum and matter in
super-strong radiation fields, Springer, the Netherlands, 2016.

\bibitem{2ccc} G. Mourou, \textquotedblright The ultrahigh-peak-power laser:
present and future\textquotedblright , \textit{Appl. Phys. B} \textbf{65},
205--211 (1997), https://doi.org/10.1007/s003400050265.

\bibitem{hhg2} M. Ferray, A. L'Huillier, X. F. Li, L. A. Lompre, G. Mainfray
and C. Manus, \textquotedblleft Multiple-harmonic conversion of 1064 nm
radiation in rare gasesn\textquotedblright , \textit{J. Phys. B} \textbf{21}%
, L31--L35 (1988), https://doi.org/10.1088/0953-4075/21/3/001.

\bibitem{att} F. Krausz and M. Ivanov, \textquotedblleft Attosecond
physics\textquotedblright , \textit{Rev. of Modern Phys.}, 81, 163--234
(2009), https://doi.org/10.1103/RevModPhys.81.163.

\bibitem{imig1} O. Smirnova, Y. Mairesse, S. Patchkovskii, N. Dudovich, D.
Villeneuve, P. Corkum, M. Yu. Ivanov, \textquotedblleft High harmonic
interferometry of multi-electron dynamics in molecules\textquotedblright , 
\textit{Nature} \textbf{460} 972--977 (2009),
https://doi.org/10.1038/nature08253.

\bibitem{imig2} S. Haessler, J. Caillat, P. Salieres, \textquotedblleft
Self-probing of molecules with high harmonic generation\textquotedblright , 
\textit{Journal of Physics B} \textbf{44} 203001(1)-203001(9) (2011),
https://doi.org/10.1088/0953-4075/44/20/203001.

\bibitem{imig3} C. G. Wahlstram, J. Larsson, A. Persson, T. Starczewski, S.
Svanberg, P. Salieres, P. Balcou, A. L'Huillier, \textquotedblleft
High-order harmonic generation in rare gases with an intense short-pulse
laser\textquotedblright , \textit{Phys. Rev. A} \textbf{48} 4709-4720
(1993), https://10.1103/physreva.48.4709.

\bibitem{sol4} Sh. Ghimire, A. D. DiChiara, E. Sistrunk, P. Agostini, L. F.
DiMauro, and D. A. Reis, \textquotedblleft Observation of high-order
harmonic generation in a bulk crystal\textquotedblright , \textit{Nature
Physics} \textbf{7}, 138-141 (2011), https://doi.org/10.1038/nphys1847.

\bibitem{sol5} B. Zaks , R. B. Liu, M. S. Sherwin, \textquotedblleft
Experimental observation of electron--hole recollisions\textquotedblright , 
\textit{Nature (London)} \textbf{483} 580-583 (2012),
https://doi.org/10.1038/nature10864.

\bibitem{sol6} O. Schubert, M. Hohenleutner, F. Langer, B. Urbanek, C.
Lange, U. Huttner, D. Golde, T. Meier, M. Kira, S. W. Koch, and R. Huber,
\textquotedblleft Sub-cycle control of terahertz high-harmonic generation by
dynamical Bloch oscillations\textquotedblright , \textit{Nature Photonics} 
\textbf{8}, 119--123 (2014), https://doi.org/10.1038/nphoton.2013.349.

\bibitem{sol7} G. Vampa, T. J. Hammond, N. Thir, B. E. Schmidt, F. Legare,
C. R. McDonald, T. Brabec, and P. B. Corkum, \textquotedblleft Linking high
harmonics from gases and solidsl\textquotedblright , \textit{Nature} \textbf{%
522}, 462-464 (2015), https://doi.org/10.1038/nature14517.

\bibitem{sol10} G. Ndabashimiye, S. Ghimire, M. Wu, D. A. Browne, K. J.
Schafer, M. B. Gaarde, and D. A. Reis \textquotedblleft Solid-state
harmonics beyond the atomic limit\textquotedblright , \textit{Nature} 
\textbf{534}, 520-523 (2016), https://doi.org/10.1038/nature17660.

\bibitem{sol11} Y. S. You, D. A. Reis , and S. Ghimire, \textquotedblleft
Anisotropic high-harmonic generation in bulk crystals\textquotedblright , 
\textit{Nature Physics} \textbf{13}, 345--349 (2017),
https://doi.org/10.1038/nphys3955.

\bibitem{sol12} H. Liu, C. Guo, G. Vampa, J. L. Zhang , T. Sarmiento, M.
Xiao, P. H. Bucksbaum , J. Vuckovic, S. Fan, and D. A. Reis,
\textquotedblleft Enhanced high-harmonic generation from an all-dielectric
metasurface\textquotedblright , \textit{Nature Physics} \textbf{14:}
1006--1010 (1918), https://doi.org/10.1038/s41567-018-0233-6.

\bibitem{H2} S. A. Mikhailov, K. Ziegler, \textquotedblleft Nonlinear EM
response of graphene: frequency multiplication and the self-consistent-field
effects\textquotedblright , \textit{J. Phys. Condens. Matter} \textbf{20},
384204(1)--384204(10) (2008),
http://dx.doi.org/10.1088/0953-8984/20/38/384204/meta.

\bibitem{Mer} H. K. Avetissian, A. K. Avetissian, G. F. Mkrtchian, K. V.
Sedrakian, \textquotedblleft Creation of particle-hole superposition states
in graphene at multiphoton resonant excitation by laser
radiation\textquotedblright , \textit{Phys. Rev. B} \textbf{85},
115443(1)--115443(10) (2012), http://dx.doi.org/10.1103/PhysRevB.85.115443.

\bibitem{Mer1} H. K. Avetissian, A. K. Avetissian, G. F. Mkrtchian, K. V.
Sedrakian, \textquotedblleft\ Multiphoton resonant excitation of Fermi-Dirac
sea in graphene at the interaction with strong laser
fields\textquotedblright , \textit{J. Nanophoton.} \textbf{6},
061702(1)-061702(9) (2012), https://doi.org/10.1117/1.JNP.6.061702.

\bibitem{H3} H. K. Avetissian, G. F. Mkrtchian, K. G. Batrakov, S. A.
Maksimenko, A. Hoffmann, \textquotedblleft Multiphoton resonant excitations
and high-harmonic generation in bilayer grapheme\textquotedblright , \textit{%
Phys. Rev. B}. \textbf{88}, 165411(1)--165411(9) (2013),
http://dx.doi.org/10.1103/PhysRevB.88.165411.

\bibitem{H4} P. Bowlan, E. Martinez-Moreno, K. Reimann, T. Elsaesser, and M.
Woerner, \textquotedblleft Ultrafast terahertz response of multilayer
graphene in the nonperturbative regime\textquotedblright , \textit{Phys.
Rev. B} \textbf{89}, 041408(1)--041408(5) (2014),
https://dx.doi.org/10.1103/PhysRevB.89.041408.

\bibitem{H6} I. Al-Naib, J. E. Sipe and M. M. Dignam, \textquotedblleft
Nonperturbative model of harmonic generation in undoped graphene in the
terahertz regime\textquotedblright , \textit{New J. Phys.} \textbf{17},
113018(1)--113018(17) (2015),
http://dx.doi.org/10.1088/1367-2630/17/11/113018.

\bibitem{H7} L. A. Chizhova, F. Libisch, J. Burgdorfer, \textquotedblleft
Nonlinear response of graphene to a few-cycle terahertz laser pulse: Role of
doping and disorder\textquotedblright , \textit{Phys. Rev. B} \textbf{94},
075412(1)--075412(10) (2016), http://dx.doi.org/10.1103/PhysRevB.94.075412.

\bibitem{H8} H. K. Avetissian, G. F. Mkrtchian, \textquotedblleft Coherent
nonlinear optical response of graphene in the quantum Hall
regime\textquotedblright , \textit{Phys. Rev. B} \textbf{94},
045419(1)--045419(7) (2016), https://dx.doi.org/10.1103/PhysRevB.94.045419.

\bibitem{H9} H. K. Avetissian, A.G. Ghazaryan, G. F. Mkrtchian, K. V.
Sedrakian, \textquotedblleft High harmonic generation in Landau-quantized
graphene subjected to a strong EM radiation\textquotedblright , \textit{J.
Nanophoton.} \textbf{11}, 016004(1)--016004(9) (2017),
http://dx.doi.org/10.1117/1.JNP.11.016004.

\bibitem{H10} L. A. Chizhova, F. Libisch, and J. Burgdorfer,
\textquotedblleft High-harmonic generation in graphene: Interband response
and the harmonic cutoff\textquotedblright , \textit{Phys. Rev. B} \textbf{95}%
, 085436(1)-- 085436(8) (2017), https://doi.org/10.1103/PhysRevB.95.085436.

\bibitem{H11} D. Dimitrovski, L. B. Madsen, T. G. Pedersen,
\textquotedblleft High-order harmonic generation from gapped
graphene\textquotedblright , \textit{Phys. Rev. B} \textbf{95},
035405(1)--035405(9) (2017), https://dx.doi.org/10.1103/PhysRevB.95.035405.

\bibitem{H12} N. Yoshikawa, T. Tamaya, K. Tanaka, \textquotedblleft
High-harmonic generation in graphene enhanced by elliptically polarized
light excitation\textquotedblright ,\textit{\ Science} \textbf{356},
736--738 (2017), http://dx.doi.org/10.1126/science.aam8861.

\bibitem{H12a} A. Golub, R. Egger, C. Muller, and S. Villalba-Chavez,
\textquotedblleft Dimensionality-driven photoproduction of massive Dirac
pairs near threshold in gapped graphene monolayers\textquotedblright , 
\textit{Phys. Rev. Lett} \textbf{124}, 110403(1)--110403(7) (2020),
https://doi.org/10.1103/PhysRevLett.124.110403.

\bibitem{H13} H. K. Avetissian, G.F. Mkrtchian, \textquotedblleft Impact of
electron-electron Coulomb interaction on the high harmonic generation
process in graphene\textquotedblright , \textit{Phys. Rev. B} \textbf{97},
115454(1)--115454(9) (2018), http://dx.doi.org/10.1103/PhysRevB.97.115454.

\bibitem{H14} A. K. Avetissian, A. G. Ghazaryan, Kh. V. Sedrakian,
\textquotedblleft Third harmonic generation in gapped bilayer
graphene\textquotedblright ,\textit{\ J. Nanophoton.} \textbf{13(3)},
036010(1)--036010(13) (2019), https://doi.org/10.1117/1.JNP.13.036010.

\bibitem{H15} A. G. Ghazaryan, Kh. V. Sedrakian, \textquotedblleft
Multiphoton cross sections of conductive electrons stimulated bremsstrahlung
in doped bilayer graphene \textquotedblright , \textit{\ J. Nanophoton.} 
\textbf{13(4)}, 046004(1)--046004(14) (2019),
https://doi.org/10.1117/1.JNP.13.046004.

\bibitem{H16} A. G. Ghazaryan, Kh. V. Sedrakian, \textquotedblleft
Microscopic nonlinear quantum theory of absorption of coherent
electromagnetic radiation in doped bilayer graphene\textquotedblright , 
\textit{J. Nanophoton.} \textbf{13(4)}, 046008(1)--046008(14) (2019),
https://doi.org/10.1117/1.JNP.13.046008.

\bibitem{H17} A. K. Avetissian, A.G. Ghazaryan, K. V. Sedrakian, and B. R.
Avchyan, \textquotedblleft Induced nonlinear cross sections of conductive
electrons scattering on the charged impurities in doped
graphene\textquotedblright , \textit{\ J. Nanophoton.} \textbf{11},
036004(1)--036004(11) (2017), https://doi.org/10.1117/1.JNP.11.036004.

\bibitem{H18} A. K. Avetissian, A.G. Ghazaryan, K. V. Sedrakian, and B. R.
Avchyan, \textquotedblleft Microscopic nonlinear quantum theory of
absorption of strong EM radiation in doped graphene\textquotedblright , 
\textit{J. Nanophoton.} \textbf{12}, 016006(1)--016006(12) (2018),
https://doi.org/10.1117/1.JNP.12.016006.

\bibitem{TMD} H. Liu, Y. Li, Y. S. You, Sh. Ghimire, T. F. Heinz, and D. A.
Reis, \textquotedblleft High-harmonic generation from an atomically thin
semiconductor\textquotedblright , \textit{Nature Physics} \textbf{13},
262--265 (2017), https://doi.org/10.1038/nphys3946.

\bibitem{TMD1} G. F. Mkrtchian, A. Knorr, and M. Selig, \textquotedblleft
Theory of second-order excitonic nonlinearities in transition metal
dichalcogenides\textquotedblright , \textit{Phys. Rev. B} \textbf{100},
125401(1)--125401(7) (2020), https://doi.org/10.1103/PhysRevB.100.125401.

\bibitem{BN} G. Le. Breton, A. Rubio, N. Tancogne-Dejean, \textquotedblleft
High-harmonic generation from few-layer hexagonal boron nitride: Evolution
from monolayer to bulk response\textquotedblright , \textit{Phys. Rev. B} 
\textbf{98}, 165308(1)--165308(7) (2018),
https://doi.org/10.1103/PhysRevB.98.165308.

\bibitem{TI} H. K. Avetissian, A. K. Avetissian, B. R. Avchyan, G. F.
Mkrtchian, \textquotedblleft Multiphoton excitation and high-harmonics
generation in topological insulator\textquotedblright , \textit{J. Phys.
Condens. Matter} \textbf{30}, 185302(1)--185302(7) (2018),
https://doi.org/10.1088/1361-648X/aab989.

\bibitem{Mer2019} H. K. Avetissian and G. F. Mkrtchian, \textquotedblleft
Higher harmonic generation by massive carriers in buckled two-dimensional
hexagonal nanostructures\textquotedblright , \textit{Phys. Rev. B} \textbf{%
99,} 085432(1)--085432(10) (2019),
https://doi.org/10.1103/PhysRevB.99.085432.

\bibitem{corcumsolid} S. Almalki, A. M. Parks, G. Bart, P. B. Corkum, T.
Brabec, and C. R. McDonald, \textquotedblleft High harmonic generation
tomography of impurities in solids: Conceptual analysis\textquotedblright , 
\textit{Phys. Rev. B} \textbf{98}, 144307(1)--144307(6) (2018),
https://doi.org/10.1103/PhysRevB.98.144307.

\bibitem{Corcum} M. Lewenstein, Ph. Balcou, Ivanov M Yu, A. L'Huillier, and
P. B. Corkum, \textquotedblleft Theory of high-harmonic generation by
low-frequency laser fields\textquotedblright , \textit{Phys. Rev. A} \textbf{%
49} 2117--2132 (1994), https://doi.org/10.1103/PhysRevA.49.2117.

\bibitem{32b} C. Cohen-Tannoudji, J. Dupont-Roc, and G. Grynberg,
\textquotedblleft Photons and atoms-Introduction to Quantum
Electrodynamics\textquotedblright, Wiley, New York, USA,1989.

\bibitem{exp29} Y. J.-Yan, \textquotedblleft Theory of excitonic high-order
sideband generation in semiconductors under a strong terahertz
field\textquotedblright , \textit{Phys. Rev. B} \textbf{78}
075204(1)-075204(8) (2008), https://doi.org/10.1103/PhysRevB.78.075204.

\bibitem{exp30} J. A. Crosse and R. B. Liu, \textquotedblleft
Quantum-coherence-induced second plateau in high-sideband
generation\textquotedblright , \textit{Phys. Rev. B} \textbf{89}
121202(1)-121202(9) (2014), https://doi.org/10.1103/PhysRevB.89.121202.

\bibitem{exp31} X . T. Xie, B. F. Zhu, R. B. Liu, \textquotedblleft Effects
of excitation frequency on high-order terahertz sideband generation in
semiconductors\textquotedblright , \textit{New J. Phys.} \textbf{15},
105015(1)-105015(10) (2013), https://doi.org/10.1088/1367-2630/15/10/105015.

\bibitem{1} K. S. Novoselov, A. K. Geim, S. V. Morozov, D. Jiang, Y. Zhang,
S. V. Dubonos, I. V. Grigorieva, and A. A. Firsov, \textquotedblleft
Electric field effect in atomically thin carbon films\textquotedblright , 
\textit{Science} \textbf{306}(5696), 666--669 (2004),
http://dx.doi.org/10.1126/science.1102896.

\bibitem{2} A. H. Castro Neto, F. Guinea, N. M. R. Peres, K. S. Novoselov,
and A. K. Geim, \textquotedblleft The electronic properties of
graphene\textquotedblright , \textit{Rev. Mod. Phys}. \textbf{81}, 109--162
(2009), http://dx.doi.org/10.1103/RevModPhys.81.109.

\bibitem{exp41} M. J. Paul , Y. C. Chang , Z. J. Thompson , A. Stickel , J.
Wardini , H. Choi , E. D. Minot, B. Hou, J. A. Nees, T. B. Norris, Y. Lee,
\textquotedblleft High-field terahertz response of
graphene\textquotedblright , \textit{New J. Phys.} \textbf{15}
085019(1)-085019(9) (2013), https://doi.org/10.1088/1367-2630/15/8/085019.

\bibitem{exp42} E. Malic, T. Winzer, E. Bobkin, A. Knorr, \textquotedblleft
Microscopic theory of absorption and ultrafast many-particle kinetics in
graphene\textquotedblright\ \textit{Phys. Rev. B} \textbf{84},
205406(1)-205406(6) (2011), https://doi.org/10.1103/PhysRevB.84.205406.

\bibitem{Avet19} H. K. Avetissian, A. K. Avetissian, B. R. Avchyan, G. F.
Mkrtchian, \textquotedblleft Wave mixing and high harmonic generation at
two-color multiphoton excitation in two-dimensional hexagonal
nanostructures\textquotedblright , \textit{Phys. Rev. B} \textbf{100},
035434(1)--035434(7) (2019), https://doi.org/10.1103/PhysRevB.100.035434.

\bibitem{exp43} H. K. Avetissian, B. R. Avchyan, G. F. Mkrtchian,
\textquotedblleft Two-color multiphoton resonant excitation of three-level
atoms\textquotedblright , \textit{Phys. Rev. A} \textbf{74}
063413(1)-063413(8) (2006), https://doi.org/10.1103/PhysRevA.74.063413.

\bibitem{exp44} H. K. Avetissian, B. R. Avchyan, G. F. Mkrtchian,
\textquotedblleft Enhanced high-order-harmonic generation and wave mixing
via two-color multiphoton excitation of atoms and
molecules\textquotedblright , \textit{Phys. Rev. A} \textbf{94}
013856(1)-013856(8) (2016), https://doi.org/10.1103/PhysRevA.94.013856.

\bibitem{1a} E. V. Castro, K. S. Novoselov, S. V. Morozov, N. M. R. Peres,
J. M. B. Lopes dos Santos, J. Nilsson, F. Guinea, A. K. Geim, and A. H.
Castro Neto, \textquotedblleft Biased bilayer graphene: semiconductor with a
gap tunable by the electric field effect\textquotedblright , \textit{Phys.
Rev. Lett}. \textbf{99}, 216802(1)--216802(4) (2007),
https://doi.org/10.1103/PhysRevLett.99.216802.

\bibitem{1aa} Y. B. Zhang, T.-T. Tang, C. Girit, Z. Hao, M. C. Martin, A.
Zettl, M. F. Crommie, Y. R. Shen, and F. Wang, \textquotedblleft Direct
observation of a widely tunable bandgap in bilayer
graphene\textquotedblright , \textit{Nature} \textbf{459}, 820--823 (2009),
https://doi.org/10.1038/nature08105.

\bibitem{9} F. Guinea, A. H. C. Neto, N. M. R. Peres, \textquotedblleft
Electronic states and Landau levels in graphene stacks\textquotedblright , 
\textit{Phys. Rev. B} \textbf{73}, 245426(1)--245426(8) (2006),
https://doi.org/10.1103/PhysRevB.73.245426.

\bibitem{99} A. G. Ghazaryan, H. H. Matevosyan, Kh. V. Sedrakian, \textquotedblleft
Second and third harmonics generation by coherent sub-THz radiation
 at induced Lifshitz transitions in gapped bilayer graphene\textquotedblright , 
arXiv preprint arXiv:2007.02724, 2020.

\bibitem{22b} E. McCann and V. I. Fal'ko, \textquotedblleft Landau-level
degeneracy and quantum Hall effect in a graphite bilayer\textquotedblright , 
\textit{Phys. Rev. Lett}. \textbf{96}, 086805(1)--086805(4) (2006),
https://doi.org/10.1103/PhysRevLett.96.086805.

\bibitem{24b} M. Koshino and T. Ando, \textquotedblleft Transport in bilayer
graphene: Calculations within a self-consistent Born
approximation\textquotedblright , \textit{Phys. Rev. B} \textbf{73},
245403(1)--245403(8) (2006), https://doi.org/10.1103/PhysRevB.73.245403.

\bibitem{26b} D. S. L. Abergel and T. Chakraborty, \textquotedblleft
Generation of valley polarized current in bilayer graphene\textquotedblright
, \textit{Appl. Phys. Lett.} \textbf{95}, 062107(1)--062107(3) (2009),
https://doi.org/10.1063/1.3205117.

\bibitem{27b} E. Suarez Morell and L. E. F. Foa Torres, \textquotedblleft
Radiation effects on the electronic properties of bilayer
graphene\textquotedblright , \textit{Phys. Rev. B} \textbf{86},
125449(1)--125449(5) (2012), https://doi.org/10.1103/PhysRevB.86.125449.9

\bibitem{28b} J. J. Dean and H. M. van Driel, \textquotedblleft Graphene and
few-layer graphite probed by second-harmonic generation: Theory and
experiment\textquotedblright , \textit{Phys. Rev. B} \textbf{82},
125411(1)--125411(10) (2010), https://doi.org/10.1103/PhysRevB.82.125411.

\bibitem{29b} S. Wu, L. Mao, A. M. Jones, W. Yao, C. Zhang, and X. Xu,
\textquotedblleft Quantum-enhanced tunable second-order optical nonlinearity
in bilayer graphene\textquotedblright ,\textit{\ Nano Lett.} \textbf{12},
2032--2036 (2012), https://doi.org/10.1021/nl300084j.

\bibitem{30b} Y. S. Ang, S. Sultan, and C. Zhang, \textquotedblleft
Nonlinear optical spectrum of bilayer graphene in the terahertz
regime\textquotedblright , \textit{Appl. Phys. Lett}. \textbf{97},
243110(1)--243110(3) (2010), https://doi.org/10.1063/1.3527934.

\bibitem{31b} N. Kumar, J. Kumar, C. Gerstenkorn, R. Wang, H.-Y. Chiu, A. L.
Smirl, and H. Zhao, \textquotedblleft Third harmonic generation in graphene
and few-layer graphite films\textquotedblright , \textit{Phys. Rev. B} 
\textbf{87}, 121406(1)--121406(5) (2013),
https://doi.org/10.1103/PhysRevB.87.121406.

\bibitem{10} M. Aoki, H. Amawashi, \textquotedblleft Dependence of band
structures on stacking and field in layered graphene\textquotedblright , 
\textit{Solid State Commun}. \textbf{142}, 123--127 (2007),
https://doi.org/10.1016/j.ssc.2007.02.013.

\bibitem{16} K. Tang, R. Qin, J. Zhou, H. Qu, J. Zheng, R. Fei, H. Li, Q.
Zheng, Z. Gao, and J. Lu, \textquotedblleft Electric-field-induced energy
gap in few-layer graphene\textquotedblright , \textit{J. Phys. Chem}. C 
\textbf{115,} 9458--9464 (2011), https://doi.org/10.1021/jp201761p.

\bibitem{HHGarxiv} H. K. Avetissian, A. K. Avetissian, A. G. Ghazaryan, G.
F. Mkrtchian, and Kh. V. Sedrakian, \textquotedblleft High-harmonic
generation at particle-hole multiphoton excitation in gapped bilayer
graphene\textquotedblright , \textit{J. Nanophoton.} \textbf{14},
026004(1)--026004(15) (2020), https://doi.org/10.1117/1.JNP.14.026004.

\bibitem{Hwang} E. H. Hwang and S. Das Sarma, \textquotedblleft Acoustic
phonon scattering limited carrier mobility in two-dimensional extrinsic
graphene\textquotedblright , \textit{Phys. Rev. B} \textbf{77}, 115449
(2008), https://doi.org/10.1103/PhysRevB.77.115449.

\bibitem{Viljas} J. K. Viljas and T. T. Heikkila, \textquotedblleft
Electron-phonon heat transfer in monolayer and bilayer
graphene\textquotedblright ,\textit{\ Phys. Rev. B} \textbf{81}, 245404
(2010), https://doi.org/10.1103/PhysRevB.81.245404.

\bibitem{cascadlaser} C. Berger, Z. Song, X. Li, X. Wu, N. Brown, C. Naud,
D. Mayou, T. Li, J. Hass, A. N. Marchenkov, E. H. Conrad, P. N. First, and
W. A. de Heer, \textquotedblleft Electronic confinement and coherence in
patterned epitaxial graphene\textquotedblright , \textit{\ Science} \textbf{%
312}(5777), 1191-1196 (2006), http://dx.doi.org/10.1126/science.1125925.

\bibitem{cascad} A. L. Friedman, J.L.Tedesco, P.M.Campbell, J.C. Culbertson,
E. Aifer, F. K. Perkins, R.L. Myers-Ward, J. K.Hite , C. R. Eddy Jr, G.
G.Jernigan, and D. K. Gaskill, \textquotedblleft Quantum linear
magnetoresistance in multilayer epitaxial graphene\textquotedblright , 
\textit{\ Nano letters} \textbf{10}(10), 3962-3965, (2010),
http://dx.doi.org/10.1021/nl101797d.

\bibitem{nonlinopt} H. K. Avetissian , B. R. Avchyan, G. F. Mkrtchian,
\textquotedblleft Efficient generation of moderately high harmonics by
multiphoton resonant excitation of atoms\textquotedblright , \textit{Phys.
Rev. A} \textbf{77} 023409(1)-023409(8) (2008),
http://dx.doi.org/10.1103/PhysRevA.77.023409.

\bibitem{laser} X. Ch. Zhang, A. Shkurinov and Y. Zhang, \textquotedblleft
Extreme terahertz science\textquotedblright , \textit{Nature Photonics} 11,
16--18 (2017), https://doi.org/10.1038/nphoton.2016.249.

\bibitem{laser1} G. Wysock, R. Lewick, R. F. Curl, F. K. Tittel, L. Diehl,
F. Capasso, M. Troccoli, G. Hoier, D. Bour, S. Corzine, R. Maulini, M.
Giovannini, J. Faist, \textquotedblleft Widely tunable mode-hop free
external cavity quantum cascade lasers for high resolution spectroscopy and
chemical sensing\textquotedblright , \textit{Applied Physics B} \textbf{92}
305--311 (2008), https://doi.org/10.1007/s00340-008-3047-x.
\end{thebibliography}
\end{document}